# Temporal Scattering at Irremovable Exceptional Points in Lossless Drude Media

Neng Wang,[1] Shuyong Chen,[2] and Guo Ping Wang[1,*]

[1]*State Key Laboratory of Radio Frequency Heterogeneous Integration, College of Physics and Optoelectronic Engineering, Shenzhen University, Shenzhen 518060, China*

[2]*College of Electronics and Information Engineering, Shenzhen University, Shenzhen 518060, China*

*Corresponding author: gpwang.szu.edu.cn

**Abstract**

We investigate temporal scattering in lossless Drude media and reveal an overlooked role of the zero-frequency flat band associated with static polarization charge. This flat band forms an exceptional line spanning all wavenumbers and can be directly excited during temporal scattering at photonic time interfaces, generating non-propagating static fields alongside the usual reflected and transmitted waves. Eigenvector coalescence at the corresponding exceptional points leads to two distinctive features absent in previously studied systems: a static mode whose amplitude increases linearly with time, and an additional static component arising from the system's generalized eigenvector. Remarkably, these effects occur without violating total energy conservation, underscoring the Hermitian nature of the dynamics. Our findings present a new physical picture of temporal scattering, sharply distinct from that in dispersionless and Lorentz-dispersive media.

## I. Introduction

In recent years, temporal metamaterials [1-3] have emerged as a groundbreaking class of media whose constitutive parameters can be modulated in real time. This capability offers unprecedented control over electromagnetic wave propagation [4-9] and enables phenomena unattainable in time-invariant systems, including photonic time crystals [10-16], magnetic-free nonreciprocity [17,18], non-Hermiticity without loss or gain [19-21], and broadband frequency conversion [22-25]. Temporal metamaterials often feature multiple sharp photonic time interfaces, where the material experiences sudden changes in its temporal properties. As a result, the primary physics governing the interaction of electromagnetic waves with these materials is embedded in temporal scattering at these time interfaces.

Xiao et al. [26] had previously implemented an investigation of temporal reflection and transmission in dispersionless media. Besides the frequency shifts, the temporal boundary conditions were also heuristically discussed. Later, the framework was extended to Lorentz-dispersive media, revealing qualitatively different boundary conditions and mode-conversion behavior compared with the dispersionless case [27-29]. This highlighted the crucial role of dispersion in temporal metamaterials [30-35]. Although several experimental techniques to achieve photonic time interfaces have been proposed [22, 23, 36], the first direct measurement of temporal reflection in electromagnetics could be dated back to the work by Moussa et al., where transmission line metamaterials in the low frequency regime were utilized [37]. More recently, nonlinear effects have been incorporated, enabling second-harmonic generation in temporal scattering [38, 39].

In this work, we investigate temporal scattering at time interfaces in Drude media, a classical model for conductive materials. We focus on the flat band at zero frequency, corresponding to the static polarization charge mode—an aspect largely overlooked in previous studies. We show that, in the lossless limit, this flat band is degenerate and forms an irremovable exceptional line (EL) spanning all wavenumbers. Temporal scattering can excite this flat band, converting part of the incident wave into non-propagating static fields in addition to the conventional reflected and transmitted waves.

Furthermore, the coalescence of eigenvectors at the exceptional points (EPs) means that the scattered field cannot be fully expanded in the eigenmodes of the lossless Drude medium. This results in two distinctive features: (1) the amplitude of the static mode grows linearly with time and (2) an additional static field component emerges, associated with the generalized eigenvector of the system. Despite these unconventional effects, total electromagnetic energy remains conserved, consistent with the Hermitian nature of the system. Our findings reveal intrinsic differences between temporal scattering in lossless Drude media and in previously studied systems [26-29, 33, 40, 41]. This work deepens the understanding of temporal scattering near singular points and opens new avenues for exploiting dispersive temporal metamaterials in applications.

## II. Full Band Dispersions of Drude Media

Let us consider an isotropic Drude medium described by the following time-dependent differential equation:

$$\partial_t^2 P_x + \gamma \partial_t P_x = \omega_p^2 E_x, \quad (1)$$

where $\omega_p$ is the plasma frequency, $\gamma$ is the damping rate, $E_x$ and $P_x$ are the electric field and polarization charge. Here, we have assumed that the electromagnetic wave is linearly polarized with the electric and magnetic fields along the x and y directions, respectively. Also, the vacuum permittivity $\varepsilon_0$ and permeability $\mu_0$ have been set as unity for the sake of simplicity. Combining the Maxwell's equations:

$$\partial_z E_x = -\partial_t H_y, \quad \partial_z H_y = -\partial_t D_x = -\partial_t E_x - \partial_t P_x, \quad (2)$$

and introducing the bound current density $J_x = \partial_t P_x$, we obtain the Schrodinger-like eigenvalue equation as

$$i\partial_t \begin{pmatrix} E_x \\ H_y \\ P_x \\ J_x \end{pmatrix} = \begin{pmatrix} 0 & -i\partial_z & 0 & -i \\ -i\partial_z & 0 & 0 & 0 \\ 0 & 0 & 0 & i \\ i\omega_p^2 & 0 & 0 & -i\gamma \end{pmatrix} \begin{pmatrix} E_x \\ H_y \\ P_x \\ J_x \end{pmatrix} = \hat{H} \cdot \vec{\psi}, \quad (3)$$

where $\hat{H}$ denotes the Hamiltonian of the system and $\vec{\psi} = (E_x, H_y, P_x, J_x)^T$ is the wave function. For a plane wave mode with a spatial dependence $e^{ikz}$, where $k$ is the wavenumber, we can replace the operator $-i\partial_z$ with $k$ in Eq. (3). The eigenvalues of $\hat{H}$ then yield the band dispersions $\omega(k)$ of the Drude medium.

The real and imaginary parts of the four bands of the Drude medium with $\gamma = 0.3\omega_p$ are shown in Figs. 1(a) and 1(b), respectively. The band $\omega_4(k)$ ($\omega_1(k)$) with a positive (negative) real part corresponds to the forward (backward) propagating mode, due to the positive $\partial \mathrm{Re}(\omega_4)/\partial k > 0$ (negative, $\partial \mathrm{Re}(\omega_1)/\partial k < 0$) group velocity. In addition to these two bands, there are two flat bands with real parts remaining zero ($\mathrm{Re}(\omega_{2,3}) = 0$). These bands are referred to as static bands, as their corresponding modes are neither oscillating nor propagating ($\partial \mathrm{Re}(\omega_{2,3})/\partial k = 0$). For a finite $\gamma$, the two static bands differ in their imaginary parts, as shown in Fig. 1(b). The imaginary part of one static band is zero regardless of $\gamma$, while that of the other is negative for positive $\gamma$, indicating that the lossy Drude medium supports both time-invariant and decaying static modes.

The static bands are typically overlooked, which is understandable since they cannot be excited in time-invariant systems where frequencies remain nonzero and conserved. However, in a time-varying system, scattering at the time interface leads to frequency shifts, allowing these static bands to become accessible and observable.

For a nonzero $\gamma$, $\hat{H}$ has four linearly independent eigenvectors, which form a complete but non-orthogonal basis. These eigenvectors correspond to the four eigenmodes of the lossy Drude medium. Thus, for an electromagnetic wave with a single wavenumber $k$ propagating inside the lossy Drude medium, the wave function can generally be expanded as

$$\vec{\psi}(t) = \sum_{i=1}^{4} a_i \vec{\varphi}_i e^{-i\omega_i t} e^{ikz}, \quad (4)$$

where $\omega_i$ and $\vec{\varphi}_i$ are the eigenvalues and corresponding eigenvectors of $\hat{H}$, and $a_i$ are the expansion coefficients. Suppose there is a time interface at $t = t_0$, after which the medium becomes a lossy Drude medium, the expansion coefficients $a_i$ can be determined from the continuity of the wave function at the time interface:

$$(a_1, a_2, a_3, a_4)^T = \hat{M}_1^{-1} \cdot \vec{\psi}(t_0^-), \quad (5)$$

Where

$$\hat{M}_1 = (\vec{\varphi}_1, \vec{\varphi}_2, \vec{\varphi}_3, \vec{\varphi}_4) \quad (6)$$

is the modal matrix of $\hat{H}$ in case of $\gamma \neq 0$, and $\vec{\psi}(t_0^-)$ is the wave function just before the time interface, characterizing the state of incident wave. According to Eq. (4), $a_4$ ($a_1$) represents the complex amplitude of the transmitted (reflected) wave, which can be used to calculate the transmission (reflection) coefficient through a suitable normalization. Furthermore, static waves are excited when $a_2$ or $a_3$ is nonzero. These static components do not belong to either the transmitted or reflected waves, as they do not propagate, as discussed previously. Consequently, the temporal scattering in the Drude medium differs significantly from that in dispersionless [26, 40, 41] and Lorentz dispersive [27-29, 33] media.

### III. General Eigenvector Analysis at Exceptional Points

The situation becomes particularly interesting when the Drude medium transitions to a lossless state. As $\gamma$ approaches zero, the imaginary parts of all four bands vanish, leading to the coalescence of the two static bands in both their real and imaginary parts. Importantly, every point on the flat band corresponds to an exceptional point (EP) other than a diabolic point. This can be verified by examining the phase rigidities of the two static bands. As shown in Fig. 1(c), for any arbitrary wavenumber $k$, the phase rigidities of the two static bands simultaneously drop to zero as $\gamma \to 0$, indicating the coalescence of their eigenvectors [42]. Therefore, the static band for $\gamma = 0$ manifests itself as an exceptional line (EL). Once $\gamma = 0$, this EL appears at $\omega = 0$ and is irremovable, remaining fixed even when the plasma frequency $\omega_p$ is varied. This has never been mentioned in previous research.

At the EP, both the number of eigenvalues and the eigenvectors of $\hat{H}$ reduce from 4 to 3. According to Eq. (3), the eigenvalues and corresponding eigenvectors of $\hat{H}$ for $\gamma = 0$ are analytically obtained as:

$$\omega_{1,2,3} = \{0, -\sqrt{k^2 + \omega_p^2}, \sqrt{k^2 + \omega_p^2}\}, (7)$$

and

$$(\vec{X}_1, \vec{Y}_1, \vec{Z}_1) = \begin{pmatrix} 0 & -i\omega_2\omega_p^{-2} & -i\omega_3\omega_p^{-2} \\ 0 & -ik\omega_p^{-2} & -ik\omega_p^{-2} \\ 1 & i\omega_2^{-1} & i\omega_3^{-1} \\ 0 & 1 & 1 \end{pmatrix}. (8)$$

According to Eqs. (7) and (8), the lossless Drude medium supports three eigenmodes. The static mode contains only the polarization charge component, while the forward- and backward-propagating modes are linearly polarized. Due to incompatible dimensions, the eigenvectors cannot form a complete basis. However, a complete basis can be constructed using the generalized eigenvectors of $\hat{H}$, also known as the canonical basis [43]. The three eigenvectors are generalized eigenvectors of rank 1, while the generalized eigenvector of rank 2, $\vec{X}_2$, is obtained as follows [43]:

$$\hat{H}(\gamma = 0) \cdot \vec{X}_2 = \vec{X}_1, (9)$$

Which yields

$$\vec{X}_2 = (0, k^{-1}, 0, -i)^T. (10)$$

The third element of $\vec{X}_2$ can be arbitrary; however, for simplicity, we set it to zero here.

To solve Eq. (3) in case 0f $\gamma = 0$, we first impose the following transformation:

$$i\partial_t \hat{M}^{-1} \cdot \vec{\psi} = \hat{M}^{-1} \cdot \hat{H} \cdot \hat{M} \cdot \hat{M}^{-1} \vec{\psi} = i\partial_t \vec{\psi}' = \hat{J} \cdot \vec{\psi}', (11)$$

Where

$$\hat{M}_2 = (\vec{X}_1, \vec{X}_2, \vec{Y}_1, \vec{Z}_1) \quad (12)$$

is the generalized modal matrix of $\hat{H}$,

$$\hat{J} = \begin{pmatrix} \omega_1 & 1 & & \\ 0 & \omega_1 & & \\ & & \omega_2 & 0 \\ & & 0 & \omega_3 \end{pmatrix} \quad (13)$$

is the Jordan normal form of $\hat{H}$ and $\vec{\psi}' = \hat{M}^{-1} \cdot \vec{\psi}$ is the wave function in the new basis. Inserting Eq. (13) into Eq. (11), we easily obtain

$$\vec{\psi}' = \begin{pmatrix} b_1 e^{-i\omega_1 t} - it b_2 e^{-i\omega_1 t} \\ b_2 e^{-i\omega_1 t} \\ b_3 e^{-i\omega_2 t} \\ b_4 e^{-i\omega_3 t} \end{pmatrix} e^{ikz}, \quad (14)$$

Where $b_i$ are expansion coefficients to be determined. Then the original wave function is given by

$$\vec{\psi} = \hat{M} \cdot \vec{\psi}' = [b_1 \vec{X}_1 + b_2 (-it \vec{X}_1 + \vec{X}_2) + b_3 \vec{Y}_1 e^{-i\omega_2 t} + b_4 \vec{Z}_1 e^{-i\omega_3 t}] e^{ikz}. \quad (15)$$

In Eq. (15), we have used $\omega_1 = 0$. Compared to Eq. (4), the degenerate eigenvector is replaced by the generalized eigenvector, and an additional term, $-ib_2 t \vec{X}_1$, is included in Eq. (15). Similarly, the expansion coefficients $b_i$ can be determined from the continuity of the wave function $\vec{\psi}$ at the time interface. It is convenient to set the time interface at $t = 0$, which eliminates the time dependence in Eq. (15). The temporal boundary conditions then become:

$$\vec{\psi}(0^-) = \hat{M}_2 \cdot \vec{\psi}'(0^+) = \hat{M}_2 \cdot (b_1, b_2, b_3, b_4)^T, \quad (16)$$

Which yields

$$(b_1, b_2, b_3, b_4)^T = \hat{M}_2^{-1} \cdot \vec{\psi}(0^-). \quad (17)$$

As per Eq. (15), the static modes could also be excited after the time interface when $\gamma = 0$, similar to the lossy case. The coefficients of static modes, $b_1$ and $b_2$, are simultaneous zero only when the incident wave function at the time interface, $\vec{\psi}(0^-)$, is a linear combination of $\vec{Y}_1$ and $\vec{Z}_1$. On the other hand, the amplitudes of the transmitted ($b_4$), reflected ($b_3$) and static waves ($b_2, b_3$) can be optimized by tailoring $\vec{\psi}(0^-)$ according to Eq. (17).

Moreover, in stark contrast to the lossy case where the static fields are time-invariant or decaying, the static fields inside the lossless Drude medium are composed of the time-invariant ($b_1\vec{X}_1 + b_2\vec{X}_2$) and time-growing ($-ib_2t\vec{X}_1$) components, as per Eq. (15). According to the expressions of $\vec{X}_1$ in Eq. (8) and $\vec{X}_2$ in Eq. (10), the static fields contain no electric field component. The static magnetic field and the bound current are purely time-invariant, while the static polarization charge includes both the time-invariant and time-growing components. Thus, the static part of polarization charge grows linearly with time from a finite value after the time interface.

### IV. Numerical Results and Discussions

In Fig. 2, we plot the time evolution of field components after the time interface formed between air and a lossless Drude medium. The results are obtained analytically using Eqs. (15)-(17) and from full-wave simulations using the Finite Difference Time Domain (FDTD) method. The circles obtained analytically align perfectly with the lines obtained from full-wave simulations, confirming the validity of the formulas. It is clearly evident that the polarization charge, represented by the blue lines and circles, increases with time, agreeing with our previous analysis.

Although the polarization charge grows with time, the electric energy density of static fields, defined as $\frac{1}{4}\text{Re}(\mathbf{E}\cdot\mathbf{E}^* + \mathbf{E}\cdot\mathbf{P}^*)$, is always zero because the static electric field remains zero, as stated previously. Therefore, the total energy is conserved, highlighting the Hermitian nature of the system when $\gamma = 0$. The time-growing behavior of the polarization charge can be understood using the Lorentz oscillator model, which is applicable to a

Drude medium by setting the resonant frequency to zero ($\omega=0$). When the resonant frequency vanishes, the restoring force becomes negligible, allowing positive and negative point charges to be separated by any distance without doing any work.

Since the static band, as an EL, spans all wavenumbers, the theory extends beyond just plane waves. By taking the Fourier transform of Eq. (15), one can compute the space-time evolution of any pulse, containing a range of wavenumbers, as it propagates through the lossless Drude medium. Figure 3 illustrates this evolution for a pulse encountering a time interface, where the plasma frequency of the lossless Drude medium is abruptly switched. From Figs. 3(a) and 3(b), it is clear that the pulse propagates forward with minimal distortion before reaching the time interface, due to the weak dispersion of the Drude medium near the central frequency. After temporal scattering at the interface, the pulse splits into propagating and non-propagating components. The forward- and backward-propagating waves separate, leaving behind static fields with a growing polarization charge, as shown in Figs. 3(c) and 3(d). In contrast, the static magnetic field remains time-invariant, while the static electric field vanishes. Consequently, by adjusting the temporal position of the time interface, one can achieve dynamic-to-static conversion, effectively freezing the fields at a desired location.

Finally, we emphasize that the features of temporal scattering, such as the time-growing behavior of the static polarization charge, do not abruptly vanish when deviating from the EP. In Fig. 4, we display the amplitudes of the polarization charge $|P_x|$ versus time after the time interface for different damping rates of the Drude medium. The medium transitions from air to Drude medium at the time interface, similar to the setup in Fig. 2. It is observed that the polarization charge grows with time and eventually reaches a saturation value inversely proportional to $\gamma$. As discussed previously, the amplitude of one static mode remains time-invariant, while the other decays when $\gamma>0$. However, since they have opposite signs, the amplitude of their sum increases as one mode decays. Therefore, the saturation is determined by the amplitude of the time-invariant static mode. As $\gamma$ approaches zero, the saturation tends to infinity, causing the polarization charge to grow continuously with time.

## V. Conclusion

In summary, we have derived the band dispersions of the Drude medium and shown that, in the lossless limit, the degenerate flat band at zero frequency forms an irremovable exceptional line (EL) that spans all wavenumbers. This flat band, previously overlooked in studies, can be observed during scattering at a time interface, after which the medium transitions to a lossless Drude medium. Our systematic study of this time scattering reveals that the presence of the flat band excites the static magnetic field, leading to broadband frequency conversion. Moreover, the time scattering exhibits a novel characteristic due to the coalescence of eigenvectors at the exceptional point (EP). By introducing generalized eigenvectors and solving the eigenvalue equation, we analytically demonstrate that the expansion coefficient of the degenerate eigenvector increases linearly with time after the scattering event, causing a continuous growth in the polarization charge. Despite this, the total electromagnetic energy remains conserved, as there is no static electric field. Although our focus here has been on the isotropic Drude medium, our theory can, in principle, be extended to more complex systems, such as highly lossy Lorentz media with higher-order EPs and ELs [44].

**Acknowledgement**: We thank Dr. Ruo-Yang Zhang in HKUT for his insightful suggestions. This work is supported by Key Project of the National Key R&D program of China (2022YFA1404500), and National Natural Science Foundation of China (NSFC) (No. 12174263 and No. 62105213).

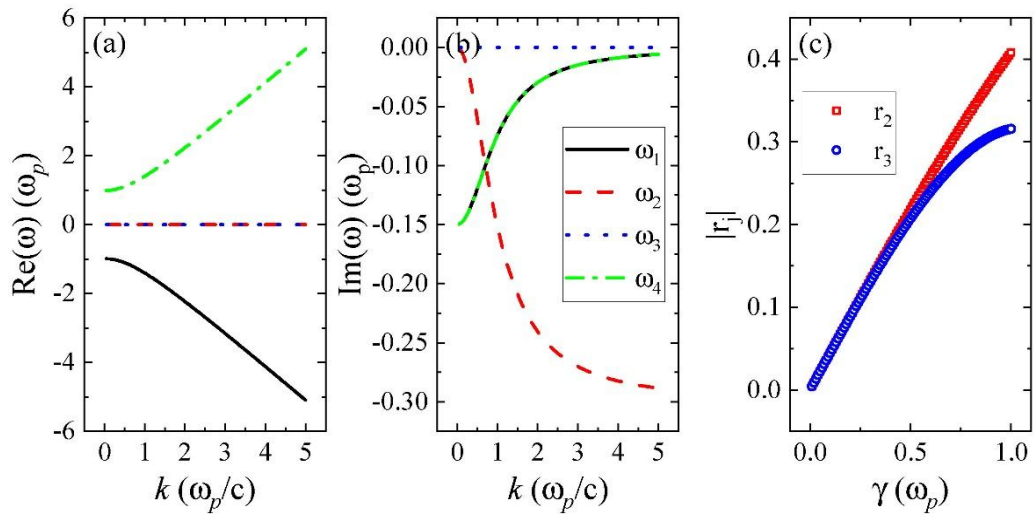

Figure 1. (a) The real and (b) the imaginary parts of the band dispersions for a Drude medium with $\gamma = 0.3\omega_p/c$. The four bands are labeled from 1 to 4 in ascending order of their real parts. (c) The phase rigidities of the two static bands versus the damping rate for $k = 0.5\omega_p/c$.

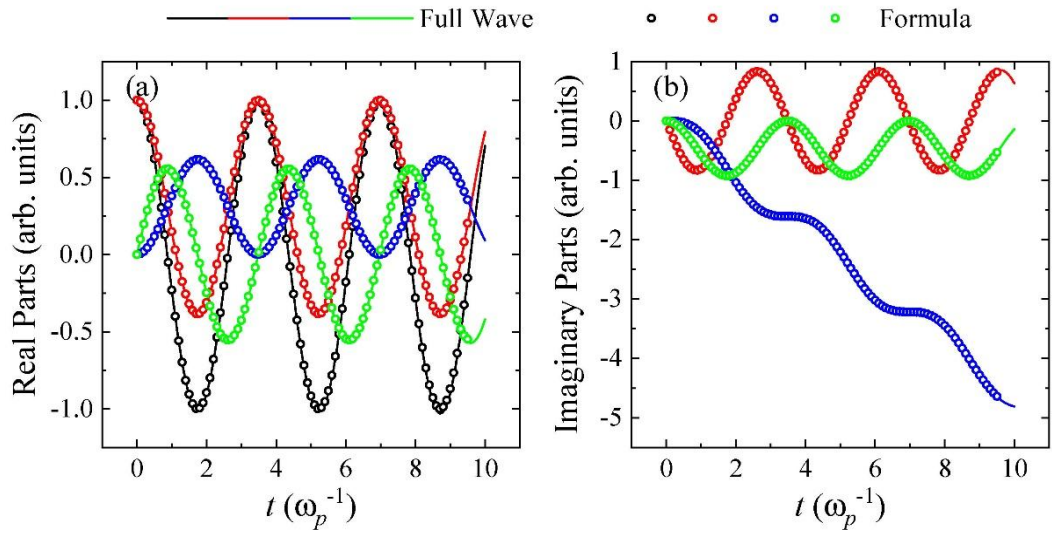

Figure 2. (a) The real and (b) the imaginary parts of the field components after the time interface for $k = 1.5\omega_p/c$. Before the time interface, the medium is air, and the wave is linearly polarized along the x-direction. At $t = 0$, the medium is abruptly switched to a lossless Drude medium. The lines and circles represent the results obtained from full wave simulations and Eq. (15), respectively, with the black, red, blue, and green curves denoting $E_x, H_y, P_x$ and $J_x$, respectively. The complex amplitudes are normalized to the electric field amplitude of the incident wave.

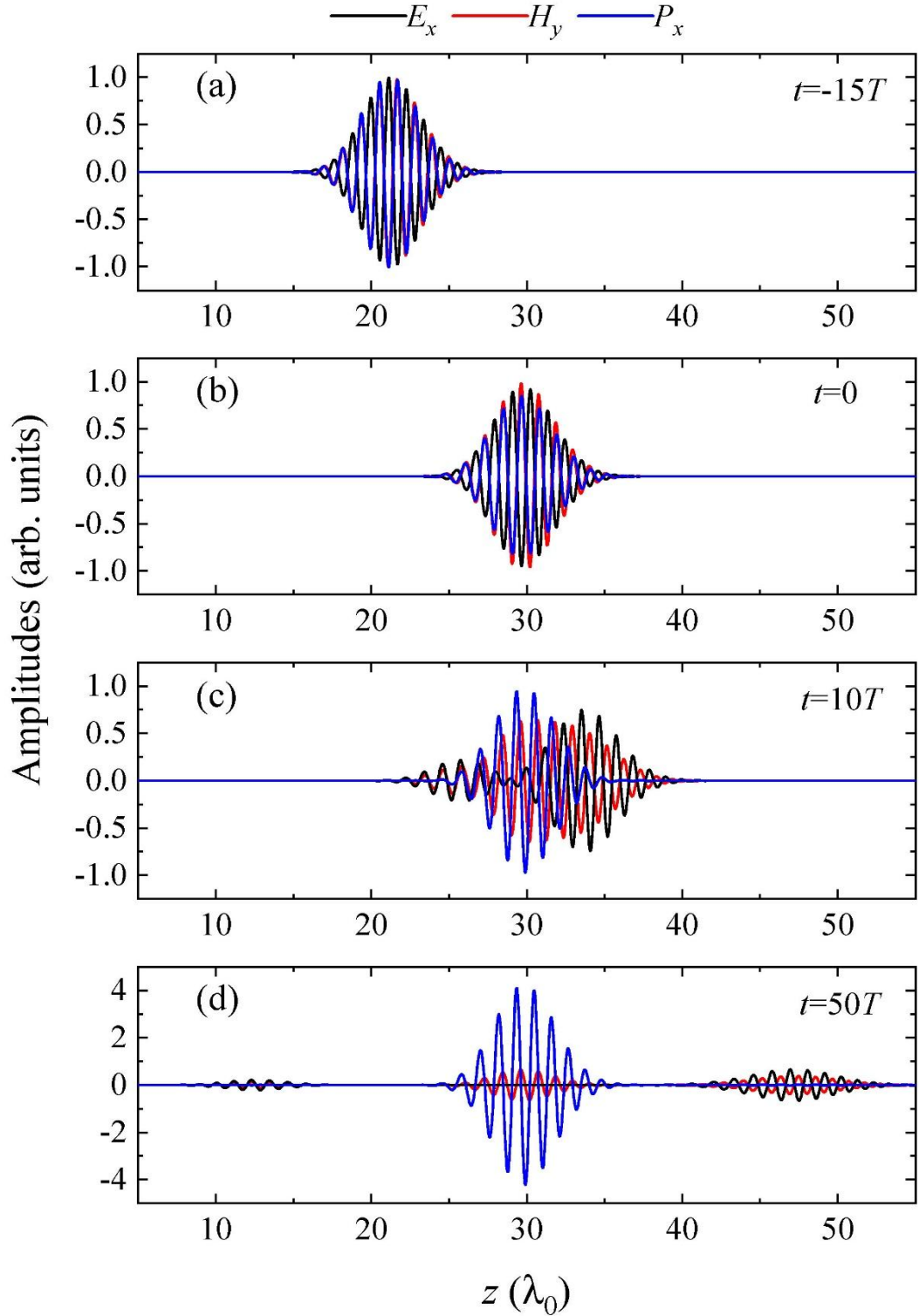

Figure 3. Spatial distributions of the electric (black), magnetic (red), and polarization charge (blue) fields at four distinct time instants. The incident pulse is given by , $\exp\left[-i\omega_c t - \frac{(t-t_c)^2}{\tau^2}\right]$ with $t_c = -35T$ and $\tau = 15T$, where $T = 2\pi/\omega_c$. For clarity, the polarization charge fields in Figs. 3(c) and 3(d) are scaled down by a factor of 100. At the time interface $t=0$, the plasma frequency of the lossless Drude medium is abruptly switched from $\omega_{p1}=0.5\omega_c$ to $\omega_{p2}=1.5\omega_c$. $\lambda_0=\omega_c/(2\pi c)$ is the vacuum wavelength at the central frequency. Amplitudes of the electric, magnetic, and polarization charge fields are normalized relative to the incident pulse's central frequency components.

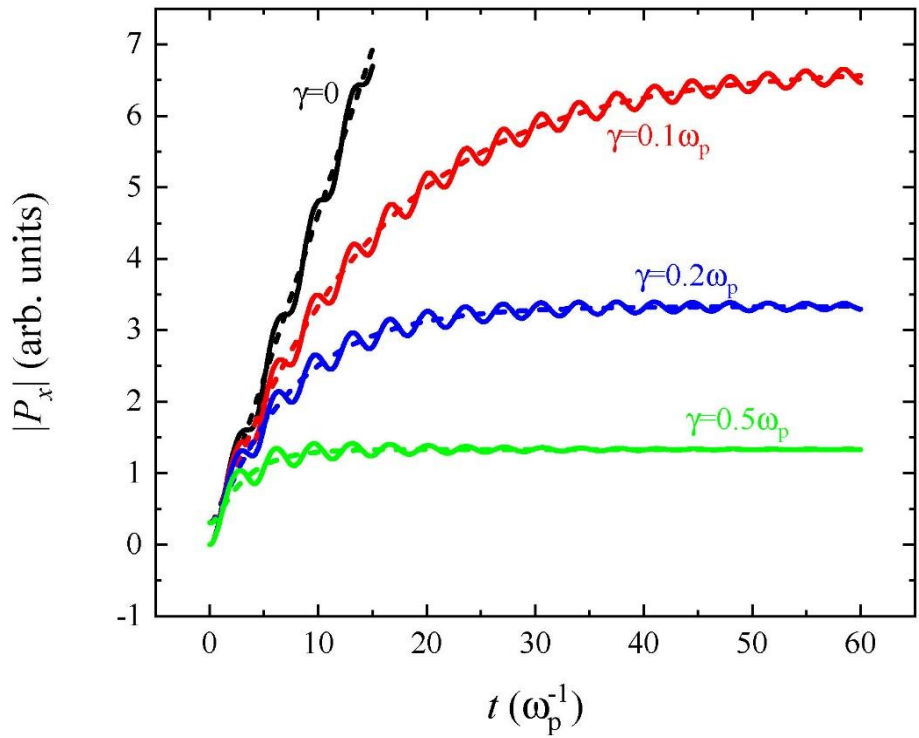

Figure 4. The absolute values of polarization charges $|P_x|$ as functions of time after the time interface ($t>0$) for different damping rates. The solid and dashed lines represent the total fields and static (non-oscillating) components, respectively. The other settings are the same as those in Fig. 2.